\documentclass[iop,twocolappendix,apj]{emulateapj}

\usepackage{apjfonts,jab,subfigure}
\shortauthors{CHEN ET AL.}
\slugcomment{}

\begin{document}
\title{Residual Energy Spectrum of Solar Wind Turbulence}
\author{C.~H.~K.~Chen\altaffilmark{1}, S.~D.~Bale\altaffilmark{1,2}, C.~S.~Salem\altaffilmark{1} and B.~A.~Maruca\altaffilmark{1}}
\affil{$^1$Space Sciences Laboratory, University of California, Berkeley, California 94720, USA; chen@ssl.berkeley.edu}
\affil{$^2$Physics Department, University of California, Berkeley, California 94720, USA}
\begin{abstract}
It has long been known that the energy in velocity and magnetic field fluctuations in the solar wind is not in equipartition. In this paper, we present an analysis of 5 years of \emph{Wind} data at 1 AU to investigate the reason for this. The residual energy (difference between energy in velocity and magnetic field fluctuations) was calculated using both the standard magnetohydrodynamic (MHD) normalization for the magnetic field and a kinetic version, which includes temperature anisotropies and drifts between particle species. It was found that with the kinetic normalization, the fluctuations are closer to equipartition, with a mean normalized residual energy of $\sigma_\mathrm{r}=-0.19$ and mean Alfv\'en ratio of $r_\mathrm{A}=0.71$. The spectrum of residual energy, in the kinetic normalization, was found to be steeper than both the velocity and magnetic field spectra, consistent with some recent MHD turbulence predictions and numerical simulations, having a spectral index close to --1.9. The local properties of residual energy and cross helicity were also investigated, showing that globally balanced intervals with small residual energy contain local patches of larger imbalance and larger residual energy at all scales, as expected for non-linear turbulent interactions.
\end{abstract}

\section{Introduction}

The solar wind displays many properties that are consistent with our current ideas of plasma turbulence \citep[see recent reviews by][]{bruno05a,horbury05,petrosyan10,matthaeus11,carbone12}. These include power spectra of field fluctuations that have power law forms \citep{coleman68}, non-Gaussian probability density functions (PDFs) of field increments \citep{marsch94} and various forms of anisotropy \citep{horbury11}. One observational feature that is less well understood, however, is that the energy in velocity and magnetic field fluctuations is not generally in equipartition. In this paper, we present new measurements to investigate the reason for this.

The power spectra of magnetic field and velocity fluctuations in the solar wind are generally interpreted in terms of magnetohydrodynamic (MHD) turbulence theory \citep{iroshnikov63,kraichnan65,goldreich95,galtier00,boldyrev06,schekochihin09} for scales larger than the ion gyroradius. Since such theories involve the non-linear interaction of Alfv\'enic wave packets, the energy in the velocity and magnetic field fluctuations has been assumed, either explicitly or implicitly, to be in equipartition, as it is for pure \citet{alfven42} waves. However, it has long been known that the energy in magnetic field fluctuations dominates the energy in velocity fluctuations \citep[e.g.,][]{belcher71,matthaeus82a,bruno85,roberts87a,tu89,marsch90a,grappin91,goldstein95c,bavassano98,podesta07a,salem09,perri10,chen11a,borovsky12a}.

Furthermore, the spectral index of the magnetic field and velocity fluctuations has been measured to be different. \citet{grappin91} showed that the velocity spectra are systematically shallower than the magnetic field spectra, which has been confirmed with more recent measurements showing an average velocity spectral index close to --3/2 and magnetic field spectral index close to --5/3 \citep{mangeney01,podesta07a,tessein09,salem09,podesta10d,chen11b,boldyrev11,borovsky12a}. This difference has also been confirmed with electric field measurements \citep{chen11b}.

An excess of magnetic energy is frequently, but not always, seen in numerical simulations of MHD turbulence, both forced and decaying, weak and strong, balanced and imbalanced and with and without a mean magnetic field \citep[e.g.,][]{oughton94,biskamp99b,beresnyak06,muller05,bigot08a,boldyrev09a,mininni09,wang11,chen11a}. The magnetic field spectrum is also often steeper than the velocity spectrum in MHD simulations \citep{muller05,mininni09,boldyrev11}.

Several possibilities have been proposed to explain these differences between the magnetic field and velocity fluctuations in the solar wind. Early explanations involved the neglect of non-MHD corrections to the Alfv\'en speed from temperature anisotropies, which would affect the normalization of the magnetic field \citep{belcher71,matthaeus82a}. Other authors \citep[e.g.,][]{hollweg87,tu89,roberts90}, however, suggested that the typical measured anisotropies were not sufficient to explain the difference and \citet{bavassano00a} showed that when incorporating the anisotropies, the magnetic fluctuations, while sometimes smaller, still dominate.

An alternative explanation is that the difference between the energy in velocity and magnetic field fluctuations, called residual energy \citep{pouquet76}, is an inherent feature of the turbulence itself. \citet{grappin83} showed numerically and analytically that turbulence in an isotropic closure theory has residual energy with a wavenumber spectrum $k^{-2}$, independent of the level of imbalance between fluxes of oppositely directed Alfv\'en wave packets. This was expanded on by \citet{muller04,muller05}, who showed, using the same closure theory and MHD simulations, that the residual energy spectrum would be $k^{-7/3}$ for an energy spectrum of $k^{-5/3}$ and $k^{-2}$ for an energy spectrum of $k^{-3/2}$. They described the presence of a small residual energy as a competition between magnetic field amplification from a local dynamo effect and equipartition from the nonlocal Alfv\'en effect \citep{kraichnan65}.

\citet{boldyrev09a} proposed that weak turbulence naturally generates a condensate of residual energy in small parallel wavenumber ($k_\|=0$) modes (which do not not have to satisfy the equipartition since they are not waves) due to mirror-invariance breaking in imbalanced turbulence \citep[see also][]{schekochihin12}. \citet{wang11} and \citet{boldyrev12c} showed analytically and numerically that this residual energy is statistically negative, i.e., the magnetic energy dominates, and has a perpendicular spectrum of $k_\perp^{-1}$. Using an anisotropic closure theory, \citet{boldyrev12a} predicted strong turbulence to have a $k_\perp^{-2}$ residual energy spectrum, independent of the particular strong turbulence model. Simulations of forced, strong reduced MHD turbulence show a scaling of $k_\perp^{-1.9}$ for both balanced and moderately imbalanced turbulence \citep{boldyrev11}. \citet{gogoberidze12b} also used an anisotropic closure theory, assuming strong, balanced, axisymmetric turbulence, finding the residual energy to be negative but with a $k^{-5/3}$ spectrum.

Further suggestions for the residual energy, that are not necessarily contradictory to the above explanations, include the presence of a separate component of convecting magnetic structures, in which there are no velocity fluctuations \citep{tu91,tu93,bruno07}. \citet{mininni09} used simulations to show that the different scaling of the velocity and magnetic fields could be due to their different amounts of intermittency, the magnetic field being more intermittent due to its ability to form thin current sheets, and therefore having a steeper power spectrum. Similarly, \citet{li11} suggested that the magnetic field takes a --3/2 spectrum to match the velocity in the absence of current sheets, although whether such discontinuities are part of the turbulence or not remains an open question \citep{borovsky08,greco08,zhdankin12a}. Compressive fluctuations have also been suggested to be a cause of the residual energy, although \citet{hollweg87} noted that their presence would lower, rather than increase, the magnetic energy.

While the amount of residual energy has been frequently measured in the solar wind, its scaling properties have been rarely documented. To our knowledge, the only measurements of residual energy scaling were done by \citet{tu89}, who found a slope near to --5/3 for an interval of fast wind measured by \emph{Helios} at relatively low frequencies (below $6\times10^{-3}$ Hz) and by \cite{podesta10d} who quoted a value of --1.75 but gave no further details. In this paper, we present a large survey of residual energy scaling measurements, compare these to the above theoretical predictions and make new measurements of the local properties of the residual energy.

\section{Data Set}
\label{sec:data}

To make reliable scaling measurements of solar wind turbulence, it is necessary to use a large amount of data. While it would be better to use a single long interval in which to measure the spectral indices, in practice this is not possible in the solar wind, so an average of the spectral indices from many shorter intervals is typically used. Histograms of inertial range spectral index measurements made in small intervals, typically a few hours long, show that they are approximately normally distributed with values ranging from --2 to --1 \citep{smith06a,vasquez07a,tessein09,chen11b,boldyrev11,borovsky12a}. Simulations also show a similar spread of values between snapshots \citep{boldyrev11} suggesting this to be an inherent feature of the turbulence. Although the mean of this distribution does not have to be the same as that calculated from a single long interval, observational \citep{borovsky12a} and simulational \citep{boldyrev11} data suggests that they are.

For the current analysis, 5 years of data from the \emph{Wind} spacecraft \citep{acuna95}, from June 2004 to April 2009, were used when the spacecraft was continuously at the first Lagrange point (L1), $\approx$230 $R_\mathrm{E}$ upstream of Earth. This period covers the declining phase of solar cycle 23 and the start of solar cycle 24. Magnetic field data from MFI \citep{lepping95}, onboard ion moments from 3DP/PESA-L \citep{lin95} and ion parameters from SWE/FC \citep{ogilvie95} were used in the analysis.

The data were split into 6 hour intervals in which various parameters, including the spectral indices, were calculated. This interval length is long enough to enable a reliable spectral index to be measured over our chosen frequency range but short enough so that there is minimal mixing of different solar wind types and the variation of the scaling with background parameters can be determined. Small data gaps in the magnetic field data and proton velocity moments were linearly interpolated over and intervals with more than 5\% data gaps or containing less than 5.4 hours of data were excluded. Several intervals in which 3DP was not in the correct mode (during January 2005, November 2006 to January 2007 and August to September 2007) were also removed, leaving 5,990 intervals of 6 hour duration.

We have chosen not to remove any discontinuities, sudden changes of the magnetic field, from the data. It is perennially debated whether these discontinuities are generated by the turbulence or are convected static structures related to processes at the Sun \citep[e.g.,][]{burlaga69,neugebauer84,horbury01b,bruno01,vasquez07b}. \citet{borovsky08} argued that the strong discontinuities are due to convected flux tubes and \citet{borovsky10} showed that they contribute significant power to the fluctuation spectrum. \citet{li11} used a few intervals free of current sheets to suggest that the magnetic field spectrum takes a --3/2 scaling in their absence, although \citet{borovsky12a} found the opposite trend, with the magnetic field spectral index becoming shallower for larger numbers of strong discontinuities. On the other hand, it has been shown that turbulence simulations can reproduce many of the properties of the solar wind discontinuities, such as the PDFs of waiting times \citep{greco08,greco09} and field rotations \citep{zhdankin12a,zhdankin12b}. \citet{borovsky10} showed that the strong discontinuities alone produce a --5/3 spectrum, which could be interpreted as them being generated by the turbulence. In this analysis, we have chosen not to remove any discontinuities from the data, since the majority are plausibly part of the turbulence.

The magnetic field, $\mathbf{B}$, from MFI and the proton velocity, $\mathbf{v}$, from 3DP/PESA-L were used to construct three further quantities: the magnetic field in velocity units, $\mathbf{b}$, and the two \citet{elsasser50} variables, $\mathbf{z}^\pm=\mathbf{v}\pm\mathbf{b}$. Since $\mathbf{z}^\pm$ correspond to Alfv\'enic wave packets traveling parallel and anti-parallel to $\mathbf{B}$ \citep{kraichnan65}, the direction of $\mathbf{B}$ was ``rectified'' based on the predicted local \citet{parker58a} spiral direction so that $\mathbf{z}^+$ corresponds to outward Alfv\'enic propagation away from the Sun and  $\mathbf{z}^-$ corresponds to inward propagation towards the Sun \citep{bruno85,roberts87a}.

The normalization of $\mathbf{b}$ is a subtle issue. Usually the normalization 
\begin{equation}
\label{eq:mhdnormalization}
\mathbf{b}=\frac{\mathbf{B}}{\sqrt{\mu_0\rho}}
\end{equation}
is used, where $\rho$ is the mass density (based on the proton density or with an estimated fraction of alpha particles). This makes the incompressible MHD equations symmetric in $\mathbf{z}^\pm$ and we call this the ``MHD normalization.'' For a kinetic plasma that is not in thermal equilibrium, such as the solar wind, the Alfv\'en wave remains noncompressive but is modified, with corrections for temperature anisotropies and relative drifts between species. One can then normalize the magnetic field as \citep{barnes79}
\begin{equation}
\label{eq:kineticnormalization}
\mathbf{b}=\frac{\mathbf{B}}{\sqrt{\mu_0\rho}}\left[1+\frac{\mu_0}{B^2}\left(p_\perp-p_\|-\sum_sm_sn_s(\Delta\mathbf{v}_s)^2\right)\right]^\frac{1}{2},
\end{equation}
where $\rho$ is the total mass density, $p_\perp$ and $p_\|$ are the total perpendicular and parallel pressures, $m_s$ is the mass of species $s$, $n_s$ is the number density of species $s$ and $\Delta \mathbf{v}_s$ is the drift of species $s$ with respect to the centre of mass velocity. We call this the ``kinetic normalization.'' In Section \ref{sec:results}, we discuss results using both normalizations.

For the MHD normalization, the mean $\rho$ for each 6 hour interval was calculated in the standard way, using proton number density data from 3DP/PESA-L and assuming that 4\% of the solar wind ions are alpha particles, the rest being protons. For the kinetic normalization, proton and alpha parameters derived from SWE measurements \citep[as described in][]{maruca11,maruca12b} were used to determine the 6 hour average values of $\rho$, $p_\perp$, $p_\|$ and $\Delta\mathbf{v}_s$. Since the electrons are lighter and generally more isotropic than the ions (due to their higher collisionality and/or kinetic instabilities \citep{stverak08}), we expect them to have a much smaller contribution in Equation \ref{eq:kineticnormalization}; they have not been included in the present analysis, but will be investigated in future work. It was also assumed that the other higher mass ions are sufficiently scarce to contribute to the normalization. An interesting property of the kinetic normalization is that it occasionally produces an imaginary magnetic field when the plasma is unstable to pressure anisotropy or drift instabilities. These intervals, along with those where no SWE data was available, were excluded from the analysis, leaving 5,396 intervals of 6 hour duration.

\begin{figure}
\epsscale{1.25}
\plotone{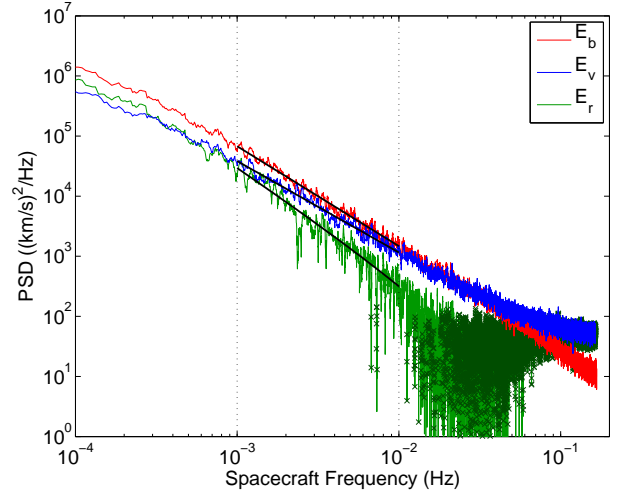}
\caption{\label{fig:spectra}Power spectra of magnetic field ($E_\mathrm{b}$), velocity ($E_\mathrm{v}$) and residual energy ($E_\mathrm{r}=E_\mathrm{v}-E_\mathrm{b}$). The light green line represents negative $E_\mathrm{r}$ and the dark green crosses represent positive $E_\mathrm{r}$. The solid black lines are fits to the measured $E_\mathrm{b}$ and $E_\mathrm{v}$ spectra and the difference of these fits and have slopes of --1.66, --1.52 and --1.96 respectively.}
\end{figure}

In each 6 hour interval, the trace power spectra of $\mathbf{v}$, $\mathbf{b}$ and $\mathbf{z}^\pm$, denoted $E_\mathrm{v}$, $E_\mathrm{b}$ and $E_\pm$, were calculated using the multitaper method with time-bandwidth product $NW=4$ \citep{percival93}. The spectra of residual energy $E_\mathrm{r}$, cross helicity $E_\mathrm{c}$, total energy $E_\mathrm{t}$, normalized residual energy $\sigma_\mathrm{r}$, normalized cross helicity $\sigma_\mathrm{c}$, Alfv\'en ratio $r_\mathrm{A}$ and Elsasser ratio $r_\mathrm{E}$, were then calculated,
\begin{eqnarray}
E_\mathrm{r}&=&E_\mathrm{v}-E_\mathrm{b},\\
E_\mathrm{c}&=&E_+-E_-,\\
E_\mathrm{t}&=&E_\mathrm{v}+E_\mathrm{b},\\
\sigma_\mathrm{r}&=&\frac{E_\mathrm{v}-E_\mathrm{b}}{E_\mathrm{v}+E_\mathrm{b}}=\frac{r_\mathrm{A}-1}{r_\mathrm{A}+1},\label{eq:sigmar}\\
\sigma_\mathrm{c}&=&\frac{E_+-E_-}{E_++E_-}=\frac{r_\mathrm{E}-1}{r_\mathrm{E}+1},\label{eq:sigmac}\\
r_\mathrm{A}&=&\frac{E_\mathrm{v}}{E_\mathrm{b}}=\frac{1+\sigma_\mathrm{r}}{1-\sigma_\mathrm{r}},\label{eq:ra}\\
r_\mathrm{E}&=&\frac{E_+}{E_-}=\frac{1+\sigma_\mathrm{c}}{1-\sigma_\mathrm{c}}.\label{eq:re}
\end{eqnarray}
While the centre of mass velocity should ideally be used to calculate these spectra, the alpha, electron and minor ion velocities are less well measured and only contribute a small fraction, so using the proton velocity for $\mathbf{v}$ is a reasonable approximation.

Example spectra of $E_\mathrm{v}$, $E_\mathrm{b}$ and $E_\mathrm{r}$, calculated over a 2 day interval (6th July 2007 00:00:00 -- 8th July 2007 00:00:00) using the kinetic normalization are shown in Figure \ref{fig:spectra}. Over most of the frequency range $E_\mathrm{b}$ is higher amplitude and steeper than $E_\mathrm{v}$. Above $\approx 3\times 10^{-2}$ Hz, $E_\mathrm{v}$ becomes larger than $E_\mathrm{b}$ due to noise in the velocity data (see Appendix). All spectral quantities were measured over the range $10^{-3}$--$10^{-2}$ Hz, which is at low enough frequencies to avoid velocity noise affecting the analysis (see Appendix), at high enough frequencies to avoid the outer scale of the turbulence and gives a large enough range to obtain reliable spectral indices. This range corresponds to typical length scales of $\sim 5\times 10^4$ km to $\sim 5\times 10^5$ km under the \citet{taylor38} hypothesis, which is smaller than the typical correlation length and larger than the typical proton gyroradius.

\begin{figure}
\epsscale{1.35}
\plotone{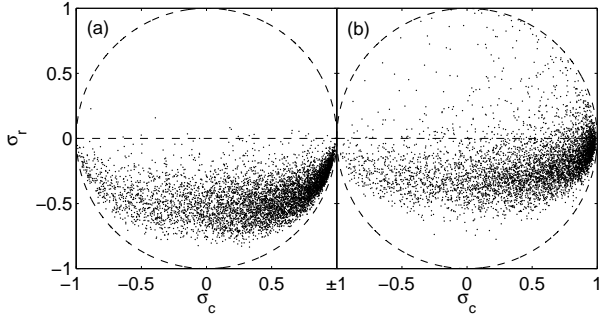}
\caption{\label{fig:sigmarsigmac}Scatter plot of normalized residual energy $\sigma_\mathrm{r}$ and normalized cross helicity $\sigma_\mathrm{c}$ using (a) MHD normalization and (b) kinetic normalization. Mathematically possible values lie within the circle.}
\end{figure}

It can be seen in Figure \ref{fig:spectra} that the residual energy spectrum is noisy, in particular above $10^{-2}$ Hz but also below, since it is the difference between two similar spectra. Several methods were attempted to measure its scaling: (a) a linear least squares fit in log space of $E_\mathrm{r}$, (b) a linear least squares fit in log space of the difference between the linear least squares fits in log space of $E_\mathrm{v}$ and $E_\mathrm{b}$, (c) a linear least squares fit in log space of $E_\mathrm{r}$ calculated using a wavelet transforms of $\mathbf{v}$ and $\mathbf{b}$. Methods (a) and (c) were found to produce a large scatter in residual energy spectral indices since they are fits to $E_\mathrm{r}$, which is noisy. While $E_\mathrm{v}$, $E_\mathrm{b}$ and $E_\mathrm{r}$ cannot, in general, all be power laws, in practice they approximately are, so method (b) was found to give the most accurate scaling of $E_\mathrm{r}$. In the rest of this analysis we use the results from method (b). Fits of $E_\mathrm{v}$ and $E_\mathrm{b}$ are marked on Figure \ref{fig:spectra} along with the difference of these fits, which can be seen to be almost power law and match $E_\mathrm{r}$ well. All other spectral indices were calculated with linear least squares fits.

In addition to the spectral indices of $E_\mathrm{v}$, $E_\mathrm{b}$, $E_+$, $E_-$, $E_\mathrm{r}$, $E_\mathrm{c}$, $E_\mathrm{t}$, and the average values of $\sigma_\mathrm{r}$, $\sigma_\mathrm{c}$, $r_\mathrm{A}$ and $r_\mathrm{E}$, three other parameters were calculated in each interval so that variation with solar wind conditions could be studied. These were the mean solar wind speed, $v_\mathrm{sw}$, the normalized fluctuation amplitude, $\delta B/B=\sqrt{\left<|\delta\mathbf{B}|^2\right>}/\left<|\mathbf{B}|\right>$, averaged over the fit range, and the proton collisional age, $A_\mathrm{c}$, obtained from the SWE parameters \citep{maruca11}.

\begin{deluxetable}{ccc}
\tablecaption{\label{tab:mhdkinetic}Mean values in the MHD and kinetic normalizations}
\tablehead{\colhead{Parameter} & \colhead{MHD} & \colhead{Kinetic}}
\startdata
$\sigma_\mathrm{r}$ & --0.43 & --0.19\\
$\sigma_\mathrm{c}$ & 0.40 & 0.46\\
$r_\mathrm{A}$ & 0.40 & 0.71\\
$r_\mathrm{E}$ & 3.37 & 4.45
\enddata
\end{deluxetable}

\section{Results}
\label{sec:results}

\subsection{MHD vs Kinetic Normalization}
\label{sec:mhdkinetic}

As described in Section \ref{sec:data}, the analysis was performed with both the standard MHD normalization of the magnetic field (Equation \ref{eq:mhdnormalization}) and the kinetic normalization with temperature anisotropies and species drifts (Equation \ref{eq:kineticnormalization}). Scatter plots of the normalized residual energy, $\sigma_\mathrm{r}$, and normalized cross helicity, $\sigma_\mathrm{c}$, (imbalance) for both normalizations are shown in Figure \ref{fig:sigmarsigmac}. The definitions of these quantities require $\sigma_\mathrm{r}^2+\sigma_\mathrm{c}^2\leq 1$ \citep{bavassano98}, i.e., the points are constrained to lie within the dashed circles.

\begin{figure}
\epsscale{1.25}
\plotone{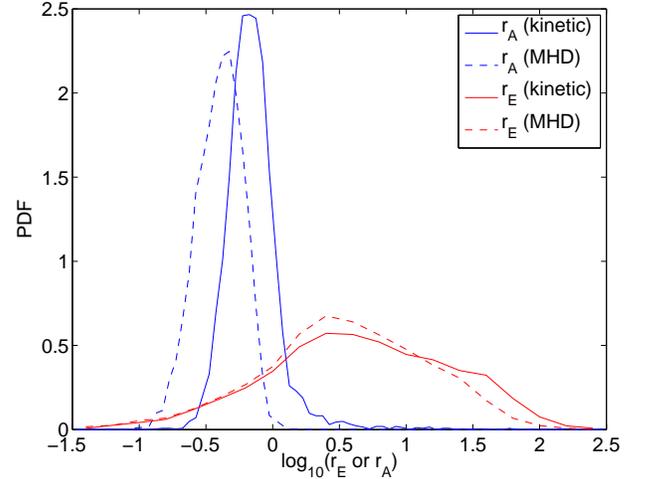}
\caption{\label{fig:rare}PDFs of Alfv\'en ratio $r_\mathrm{A}$ and Elsasser ratio $r_\mathrm{E}$ using the MHD normalization (dashed lines) and kinetic normalization (solid lines).}
\end{figure}

It can be seen that the majority of points ($\approx$ 80\% in both cases) have positive values of $\sigma_\mathrm{c}$, corresponding to dominant propagation away from the Sun, as is well known \citep[e.g.,][]{belcher71}. Periods of $\sigma_\mathrm{c}<0$ may arise naturally from the turbulence or could be due to incorrect rectification due to polarity inversions \citep{balogh99}. The mean values of $\sigma_\mathrm{c}$ are 0.40 with the MHD normalization and 0.46 with the kinetic normalization. A more significant difference can be seen in the $\sigma_\mathrm{r}$ distribution. $\sigma_\mathrm{r}$ is systematically more negative in the MHD normalization compared to the kinetic normalization, having mean values of --0.43 and --0.19 respectively. While the residual energy is still, on average, negative, as predicted by turbulence theory \citep{wang11,boldyrev12c,gogoberidze12b}, using the kinetic normalization more appropriate to the solar wind results in significantly less residual energy. 

The more natural variables in which to examine this difference, perhaps, are the Alfv\'en and Elsasser ratios, directly related to $\sigma_\mathrm{r}$ and $\sigma_\mathrm{c}$ through Equations \ref{eq:sigmar}--\ref{eq:re}. From Figure \ref{fig:rare} it can be seen that $r_\mathrm{A}$ and $r_\mathrm{E}$ are approximately log-normally distributed and there is a much broader distribution of $r_\mathrm{E}$ than $r_\mathrm{A}$. Both are more positive with the kinetic normalization, and their mean values, calculated in log space, are $r_\mathrm{A}=0.40$ and $r_\mathrm{E}=3.37$ in the MHD normalization and $r_\mathrm{A}=0.71$ and $r_\mathrm{E}=4.45$ in the kinetic normalization. These values are summarized in Table \ref{tab:mhdkinetic} and the standard error of the mean for all of these quantities is between 0.5\% and 2\%. This shows that $r_\mathrm{A}$ in the solar wind is significantly closer to unity than previous measurements suggested, but it is still $<1$ at 1 AU.

It should be pointed out that a small part of the difference between the MHD and kinetic normalizations is due to the different instruments used to measure particle density (3DP/PESA-L for the MHD normalization and SWE/FC for the kinetic normalization), with a measured, rather than estimated, alpha particle density in the kinetic normalization. This was done so that the MHD normalization can be compared to previous measurements. Calculating the results using the SWE/FC densities in the MHD normalization, however, shows that the majority of the difference is due to the temperature anisotropy and drift corrections. For the rest of this section, the kinetic normalization is used, since we consider the SWE/FC densities to be more accurate and the kinetic normalization more appropriate to the solar wind.

\subsection{Spectral Index Distributions}

\begin{figure}
\epsscale{1.25}
\plotone{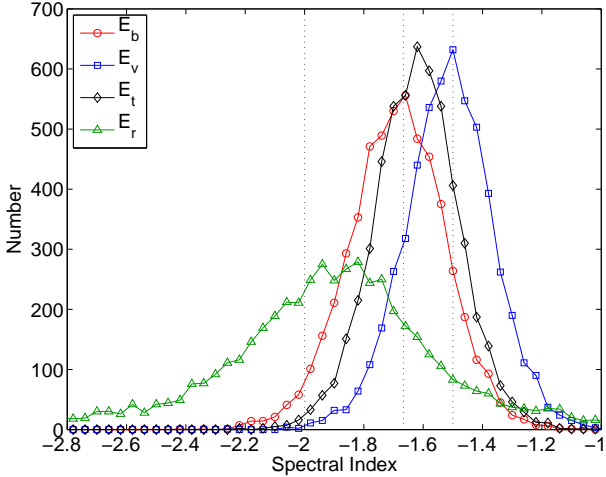}
\caption{\label{fig:histograms}Histograms of the measured spectral indices in the spacecraft frequency range $10^{-3}$--$10^{-2}$ Hz. The black dotted lines correspond to various spectral index predictions.}
\end{figure}

\begin{figure*}
\epsscale{1.1}
\plotone{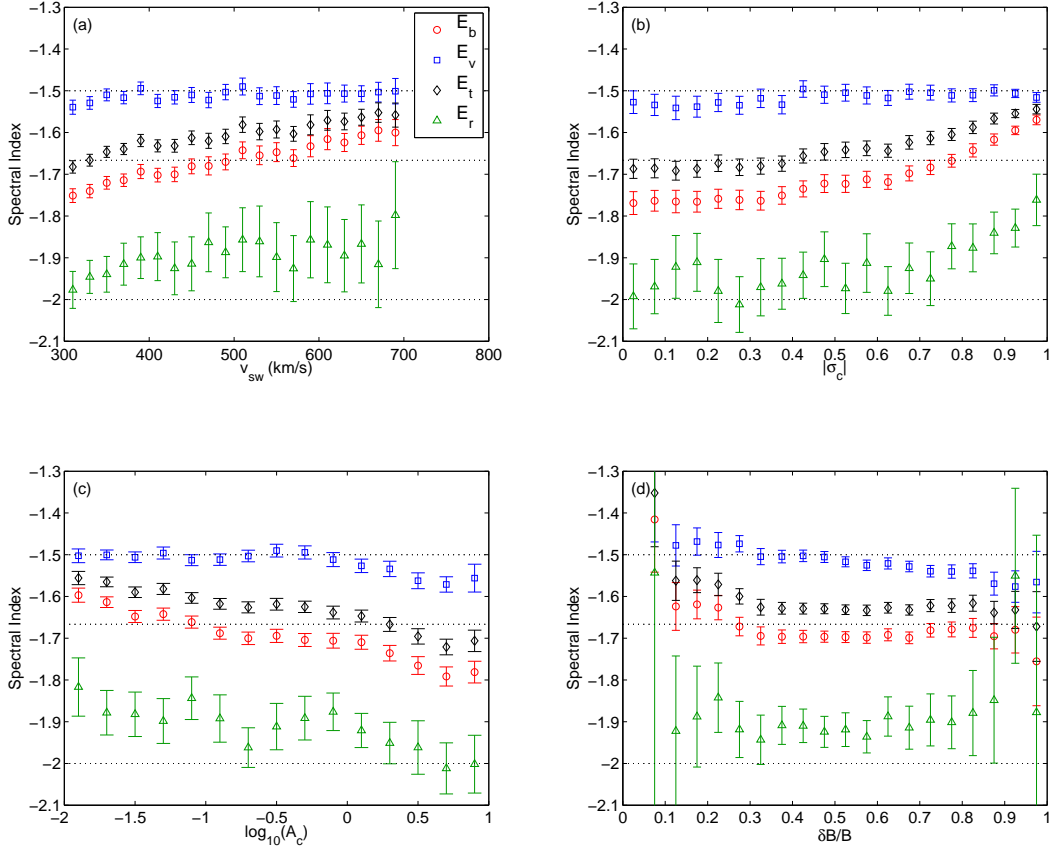}
\caption{\label{fig:parameters}Dependence of spectral indices on (a) solar wind speed $v_\mathrm{sw}$, (b) magnitude of imbalance $|\sigma_\mathrm{c}|$, (c) collisional age $A_\mathrm{c}$ and (d) fluctuation amplitude $\delta B/B$. The error bars represent 2 standard errors of the mean. The black dotted lines correspond to different spectral index predictions.}
\end{figure*}

Histograms of the spectral indices of $E_\mathrm{b}$, $E_\mathrm{v}$, $E_\mathrm{t}$ and $E_\mathrm{r}$ are shown in Figure \ref{fig:histograms}. The distributions of $E_\mathrm{b}$, $E_\mathrm{v}$ and $E_\mathrm{t}$ indices are similar to previous studies \citep{smith06a,vasquez07a,tessein09,chen11b,boldyrev11,borovsky12a}, with mean values --1.69, --1.52 and --1.62 respectively. Due to the large number of intervals, the standard error of the mean is small, 0.002 in each case. The spread of the $E_\mathrm{r}$ spectral indices is larger, which reflects the fact that this quantity is noisier since it is the difference between two similar spectra. It can still be seen, however, that the mean value of --1.91 is significantly steeper than the other three spectra, as predicted by some residual energy theories \citep{muller04,muller05,boldyrev12a}. The $E_+$, $E_-$ and $E_\mathrm{c}$ spectra (not shown) are similar to that of the total energy, having mean values of --1.62, --1.61 and --1.60 respectively. The linear correlation coefficient between the $E_\mathrm{b}$ and $E_\mathrm{r}$ spectral indices is $0.44\pm0.03$ and between the $E_\mathrm{v}$ and $E_\mathrm{r}$ spectral indices is $0.03\pm0.03$, where the uncertainty is the 99\% confidence interval. This suggests that the properties of the residual energy are related to the magnetic field, rather than the velocity.

\subsection{Dependence on Parameters}

The large data set enables the variation of the spectral indices with solar wind parameters to be studied. Figure \ref{fig:parameters} shows the mean spectral indices as a function of solar wind speed $v_\mathrm{sw}$, magnitude of imbalance $|\sigma_\mathrm{c}|$, collisional age $A_\mathrm{c}$ and fluctuation amplitude $\delta B/B$. The error bars are $\pm$ 2 standard errors of the mean. It can be seen that the velocity spectral index remains close to --3/2 for for all values of these parameters, but there are systematic trends in the other three. The magnetic field and total energy follow each other as expected, since the magnetic energy generally dominates. The residual energy also generally follows the trends of $E_\mathrm{b}$ and $E_\mathrm{t}$ but is overall steeper. The spectral indices of $E_+$, $E_-$ and $E_\mathrm{c}$ are the same to within errors as the total energy $E_\mathrm{t}$ in nearly cases, so are not shown. The only exception is that $E_-$ becomes shallower than the other spectra for large $|\sigma_\mathrm{c}|$, large $v_\mathrm{sw}$ and small $A_\mathrm{c}$. This is an artifact due to $E_-$ being low amplitude and close to the noise floor at large imbalance, which occurs in fast, low collisionality wind.

The variation of spectral indices with $v_\mathrm{sw}$ and $|\sigma_\mathrm{c}|$ appears similar, which is likely due to the underlying positive correlation between solar wind speed and imbalance. Examining the $E_\mathrm{b}$ spectral index as a function of both $v_\mathrm{sw}$ and $|\sigma_\mathrm{c}|$ indicates that $\sigma_\mathrm{c}$ is the main cause of this variation. \citet{podesta10d} also concluded that $\sigma_\mathrm{c}$ is the parameter controlling the variation of total energy spectrum. A similar explanation can be used to describe the $A_\mathrm{c}$ dependence, since $A_\mathrm{c}\propto v_\mathrm{sw}^{-1}$. Again, examining the $E_\mathrm{b}$ spectral index as a function of $A_\mathrm{c}$ and $|\sigma_\mathrm{c}|$ shows $\sigma_\mathrm{c}$ to be the main cause of this variation. It makes sense that $\sigma_\mathrm{c}$, rather than $A_\mathrm{c}$, controls the spectrum, since Alfv\'enic turbulence is not thought to depend on the collisionality \citep{schekochihin09}. The question then becomes what is causing the $|\sigma_\mathrm{c}|$ trend. At $|\sigma_\mathrm{c}|=1$, the residual energy is constrained to be $\sigma_\mathrm{r}=0$ so $E_\mathrm{v}$ and $E_\mathrm{b}$ take the same scaling, which can be seen in Figure \ref{fig:parameters}b. As $|\sigma_\mathrm{c}|$ decreases, the residual energy becomes non-zero $\sigma_\mathrm{r}<0$ and $E_\mathrm{b}$ becomes steeper than $E_\mathrm{v}$, which is consistent with the properties of the residual energy \citep{muller04,muller05,boldyrev12a,gogoberidze12b}.

The residual energy has a spectral index between --2 and --1.9 for $0<|\sigma_\mathrm{c}|<0.75$ (Figure \ref{fig:parameters}b). This is consistent with the predicted value of --2 \citep{boldyrev12a} and recent simulations of strong MHD turbulence, which cover a similar range of imbalance and have a --1.9 spectrum \citep{boldyrev11}. This scaling is not consistent with the theory of \citet{muller04,muller05}, for which the prediction is --7/3 (since the total energy spectrum is --5/3 over this range of $|\sigma_\mathrm{c}|$), or the theory of \citet{gogoberidze12b}, for which the prediction is --5/3. For $|\sigma_\mathrm{c}|>0.75$ the residual energy spectrum becomes shallower. One possibility for this is that the highly imbalanced turbulence is less evolved due to its weaker nonlinear interaction and the residual energy has not had time to develop its steady state spectrum. A more mundane, but perhaps more likely, explanation is due to inaccuracies in the normalization. At large $|\sigma_\mathrm{c}|$ the residual energy is small and the $E_\mathrm{b}$ and $E_\mathrm{v}$ spectra should be of similar amplitude, so a small error in the magnetic field normalization will cause a large error in the residual energy spectrum, biased towards shallower spectra.

The variation of spectral indices with $\delta B/B$ (Figure \ref{fig:parameters}d) shows a slight trend towards shallower spectra at low amplitudes. Examining the $E_\mathrm{b}$ spectral index as a function of both $\delta B/B$ and $|\sigma_\mathrm{c}|$ shows that both variables separately correlate to the spectral index. While the amplitudes are lower at small $\delta B/B$, the magnetic field spectra are well above the magnetometer noise floor, so the trend in $E_\mathrm{b}$ appears to be physical. This is consistent with the result of \citet{li11}, in that lower amplitude turbulence would be expected to produce fewer strong discontinuities.

\subsection{Local Properties}

\begin{figure*}
\epsscale{0.55}
\begin{center}
\subfigure{\plotone{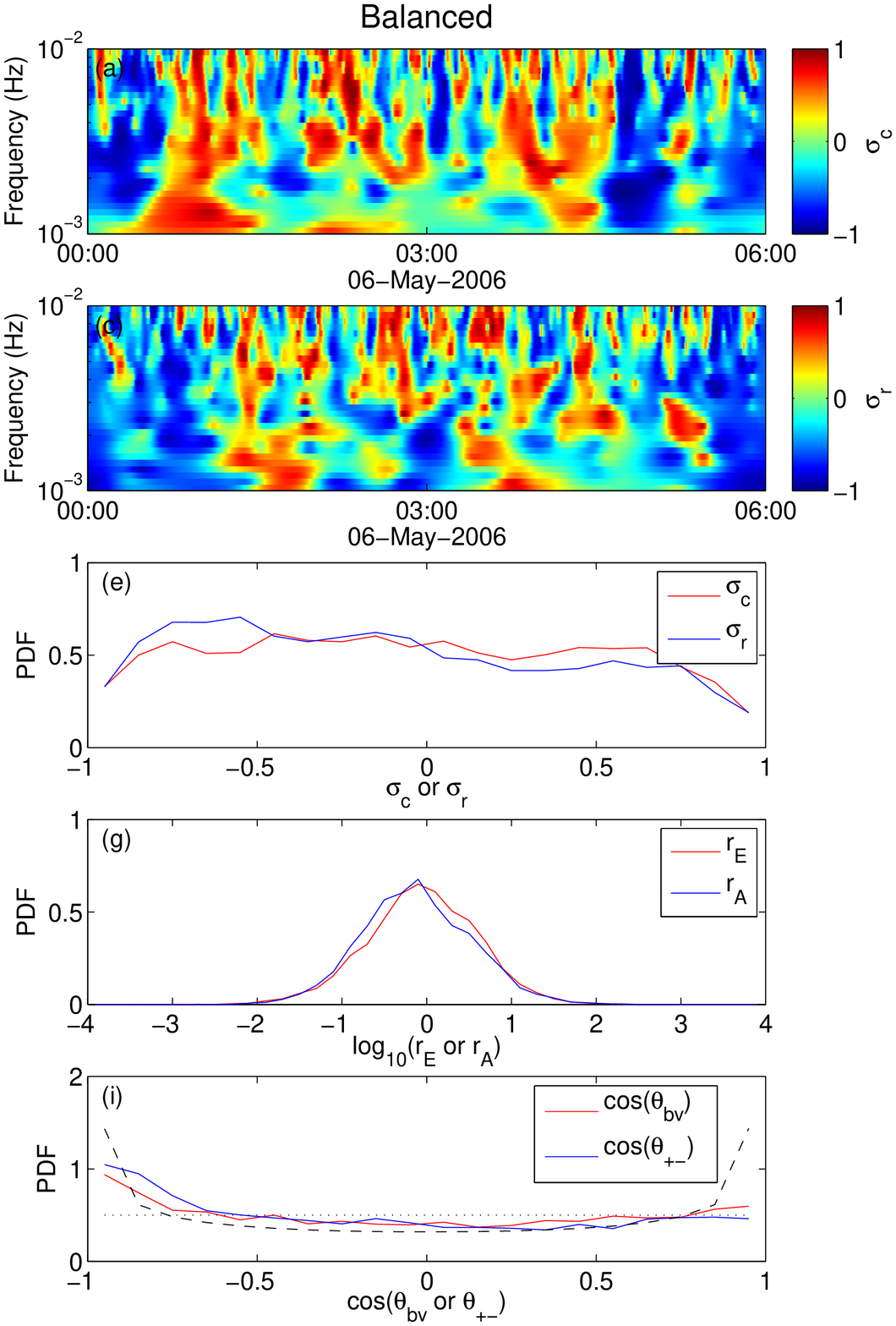}}
\subfigure{\plotone{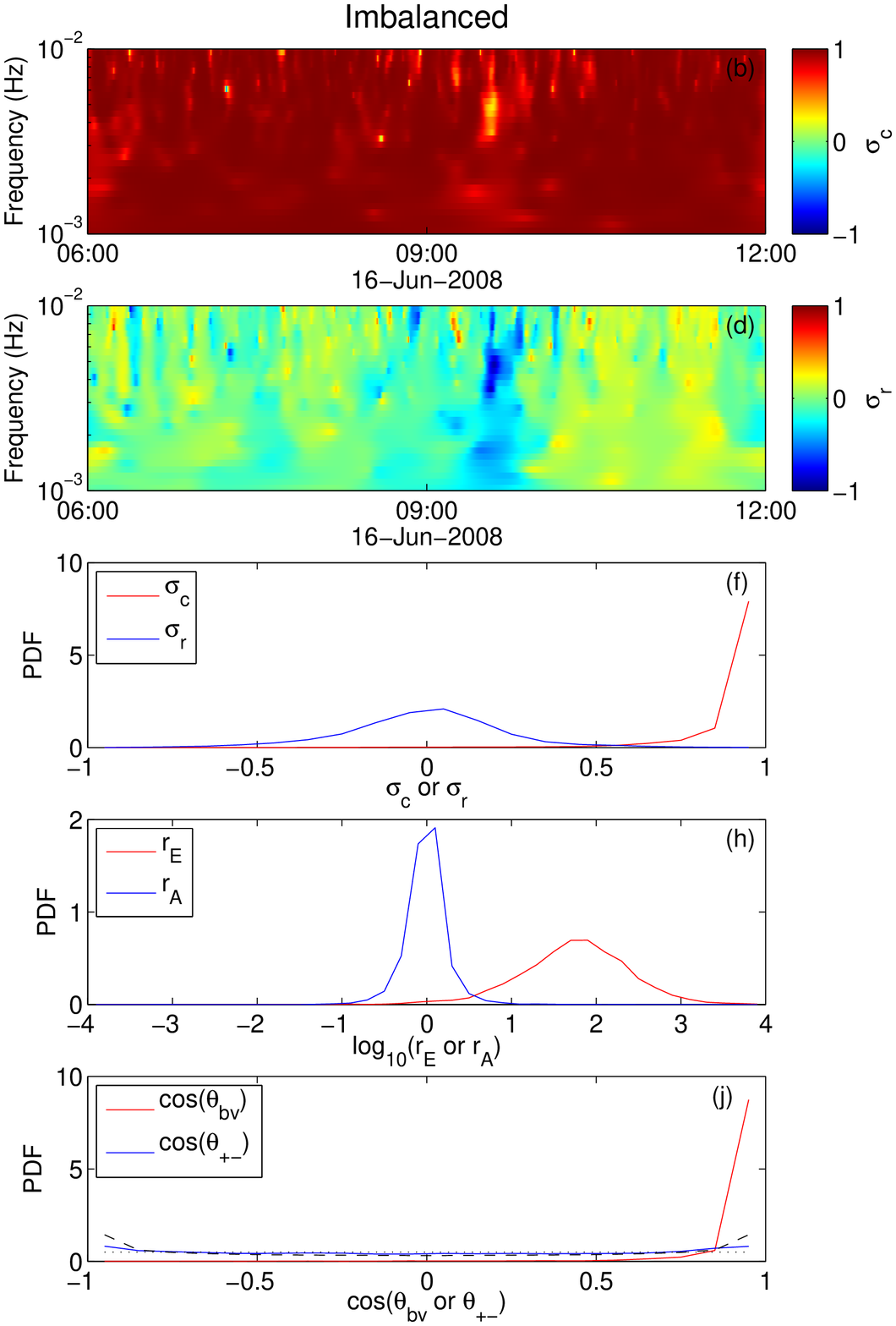}}
\caption{\label{fig:local}Local properties of globally balanced (left panels) and globally imbalanced (right panels) turbulence. (a-d) Wavelet spectrograms of $\sigma_\mathrm{c}$ and $\sigma_\mathrm{r}$. (e-j) PDFs of $\sigma_\mathrm{c}$, $\sigma_\mathrm{r}$, $r_\mathrm{E}$, $r_\mathrm{A}$, $\cos(\theta_{bv})$ and $\cos(\theta_{+-})$ at $3\times 10^{-3}$ Hz. The turbulence consists of local patches of imbalance and non-zero residual energy.}
\end{center}
\end{figure*}

Figure \ref{fig:local}a-d shows the continuous wavelet spectrograms \citep{torrence98} of normalized cross helicity $\sigma_\mathrm{c}$ and normalized residual energy $\sigma_\mathrm{r}$ for two of the intervals. Such spectrograms have been used previously to study the nature of large scale fluctuations \citep{lucek98} and flux ropes \citep{telloni12} in the solar wind. The two intervals were chosen to be globally balanced ($\sigma_\mathrm{c}=-0.13$, $r_\mathrm{E}=0.81$) and globally imbalanced ($\sigma_\mathrm{c}=0.98$, $r_\mathrm{E}=95$), covering the range of typical intervals.

It can be seen in Figure \ref{fig:local}a that the globally balanced interval is made up locally imbalanced patches at all scales as expected for Alfv\'enic turbulence \citep[e.g.,][]{matthaeus08a,perez09,boldyrev09b,perez10b}. Such self-organization could be due to the nonlinear interaction in imbalanced patches being reduced so that they persist longer \citep{perez12} or to a rapid alignment process \citep{matthaeus08a}. This interval also has local patches of both positive and negative residual energy (Figure \ref{fig:local}c), which is consistent with MHD turbulence, since the prediction of negative residual energy is for a statistical ensemble only, not individual fluctuations \citep{wang11}. This figure shows that solar wind turbulence does contain localized patches where both velocity and magnetic energy dominate, while having a globally negative residual energy ($\sigma_\mathrm{r}=-0.20$, $r_\mathrm{A}=0.70$). This can also be seen in MHD turbulence simulations \citep[][Figure 2]{perez09}, in which there are regions where either the magnetic field or velocity dominates. As expected, the imbalanced interval consists of large $\sigma_\mathrm{c}$ values (Figure \ref{fig:local}b), which limits the spread of $\sigma_\mathrm{r}$ values (Figure \ref{fig:local}d), due to the relation $\sigma_\mathrm{r}^2+\sigma_\mathrm{c}^2\leq 1$ \citep{bavassano98}.

It is important to determine whether the local patches in Figure \ref{fig:local} are physical or due to instrumental noise. It can be seen from Figure \ref{fig:spectra} that the velocity noise is $\sim 60$ km$^2$s$^{-2}$Hz$^{-1}$, whereas the total energy spectrum, over the range $10^{-3}$--$10^{-2}$ Hz, varies from $\sim3\times 10^3-1\times 10^5$ km$^2$s$^{-2}$Hz$^{-1}$, which is a factor of $\sim40$--2000 larger than the noise. The fluctuations in Figure \ref{fig:local} are on average around half the total energy, so cannot be due to this noise.

Also shown in Figure \ref{fig:local}e-j for each interval are the PDFs of $\sigma_\mathrm{c}$, $\sigma_\mathrm{r}$, $r_\mathrm{E}$, $r_\mathrm{A}$ and alignment angles $\cos(\theta_{+-})=\delta\mathbf{z}^+\cdot\delta\mathbf{z}^-/(|\delta\mathbf{z}^+||\delta\mathbf{z}^-|)$ and $\cos(\theta_{vb})=\delta\mathbf{v}\cdot\delta\mathbf{b}/(|\delta\mathbf{v}||\delta\mathbf{b}|)$, calculated from 2-point differences: $\delta\mathbf{x}=\mathbf{x}(t)-\mathbf{x}(t+\tau)$, where $1/\tau=3\times10^{-3}$ Hz. The $\sigma_\mathrm{c}$ and $\sigma_\mathrm{r}$ distributions (Figure \ref{fig:local}e,f) confirm the findings from the wavelet spectrograms, that the balanced interval is made up of patches of local imbalance and non-zero residual energy, and the imbalanced interval has a strongly peaked $\sigma_\mathrm{c}$ distribution and narrower $\sigma_\mathrm{r}$ distribution. The PDFs of $r_\mathrm{A}$ and $r_\mathrm{E}$ (Figure \ref{fig:local}g,h) are approximately log-normally distributed, with standard deviations of 0.63 for $r_\mathrm{E}$ and 0.65 for $r_\mathrm{A}$ in the balanced interval and 0.60 for $r_\mathrm{E}$ and 0.17 for $r_\mathrm{A}$ in the imbalanced interval. This shows that imbalanced turbulence has a similar spread of patches of local imbalance as balanced turbulence, but the mean value is larger, so the spread of residual energy patches is smaller.

For comparison to previous studies \citep{matthaeus08a,podesta09e,perez09,osman11c}, the PDFs of the cosine of the alignment angles are also shown (Figure \ref{fig:local}i,j). The black dashed lines show the PDF for an even distribution of angles $\theta$ and the black dotted lines show the PDF of the angle between two randomly oriented vectors. The behavior of $\cos(\theta_{bv})$, which is related to the imbalance, is roughly consistent with previous measurements \citep{matthaeus08a,podesta09e,osman11c}, having a broad distribution for small imbalance and becoming strongly peaked for large imbalance. $\cos(\theta_{+-})$, which is related to the residual energy, has a similar distribution to $\cos(\theta_{bv})$ in the balanced interval, showing that the turbulence also consists of a range of alignment angles of the Elsasser variable fluctuations.

\section{Discussion}
\label{sec:discussion}

In this paper, we have shown that when the magnetic field is converted to velocity units using a kinetic normalization that includes pressure anisotropies and drifts, the ratio of energy in velocity fluctuations to magnetic fluctuations is closer to unity, with a mean value of $r_\mathrm{A}=0.71$, compared to $r_\mathrm{A}=0.40$ with the standard MHD normalization. The residual energy, i.e., the difference in energy between velocity and magnetic field fluctuations, is statistically negative (although contains local patches of both signs at all scales) and has an average spectral index of --1.91. These properties are consistent with theoretical descriptions and numerical simulations of MHD turbulence \citep{grappin83,muller04,muller05,boldyrev09a,perez09,wang11,boldyrev11,boldyrev12a,boldyrev12c,gogoberidze12b}.

The magnetic field has an average spectral index close to --5/3, consistent with previous large studies \citep{smith06a,vasquez07a,tessein09,podesta10d,chen11b,boldyrev11,borovsky12a}. The velocity spectral index was found to be close to --3/2, as in previous studies with the \emph{Wind} \citep{mangeney01,podesta07a,salem09,podesta10d,boldyrev11} and \emph{ARTEMIS} \citep{chen11b} spacecraft. \emph{ACE} measurements, however, produce a shallower spectrum, closer to --1.4 \citep{tessein09,boldyrev11,borovsky12a}. The reason for this difference is not known, but could be related to the lower frequency range used in the studies of  \citet{boldyrev11} and \citet{borovsky12a} being affected by the outer scale and to high frequency noise (see Appendix) in the study of \citet{tessein09}. The average residual energy spectral index of --1.91 found in this paper is steeper than the two values quoted in earlier studies \citep{tu89,podesta10d}. This is likely due to our use of the kinetic normalization more appropriate to the solar wind (Section \ref{sec:mhdkinetic}); with the MHD normalization, the mean residual energy spectral index with this data set is $-1.79$.

The residual energy in the solar wind fluctuations can be injected by the driving or can arise naturally from the turbulent interactions. An interesting area of future study would be to determine how much is due to the driving and how much is due to the turbulent dynamics, perhaps by a detailed comparison to simulations of MHD turbulence.

\acknowledgments
This work was supported by NASA contract NNN06AA01C and NASA grant NNX09AE41G. \emph{Wind} data were obtained from CDAWeb ({http://cdaweb.gsfc.nasa.gov}). We thank S.~Boldyrev, T.~S.~Horbury, J.~C.~Perez, M.~P.~Pulupa, and A.~A.~Schekochihin for useful discussions.

\appendix

\section{Instrumental Noise}

\begin{figure}
\epsscale{1.15}
\plotone{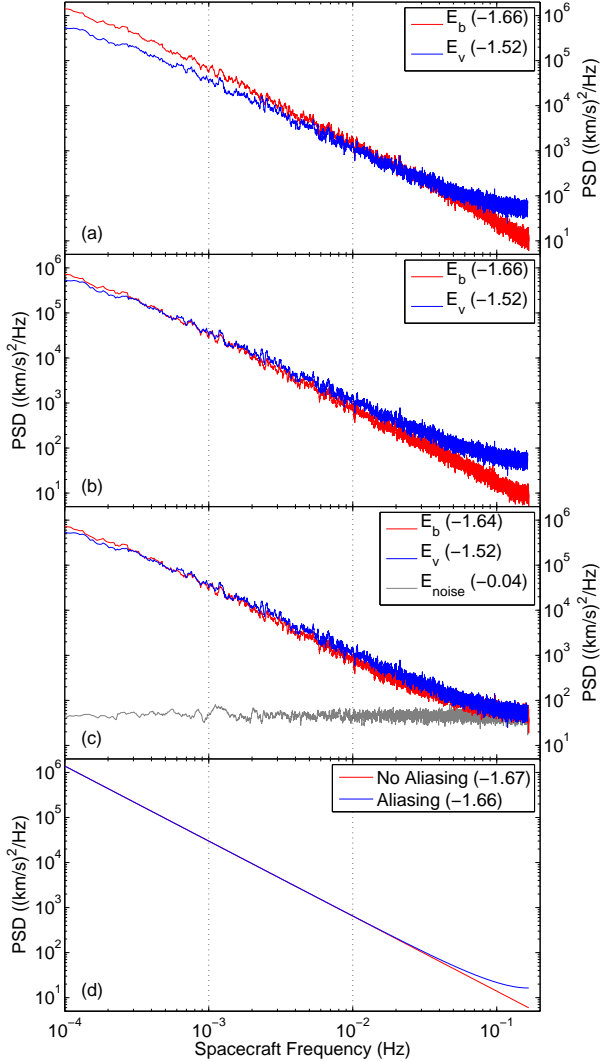}
\caption{\label{fig:noise} (a) Power spectra of magnetic field $E_\mathrm{b}$ and velocity $E_\mathrm{v}$ from Figure \ref{fig:spectra} , (b) quantization noise added to $E_\mathrm{b}$, (c) white noise added to $E_\mathrm{b}$, (d) model --5/3 spectrum with and without aliasing. Spectral indices measured between $10^{-3}$--$10^{-2}$ Hz are given in the legends.}
\end{figure}

To measure accurate spectra, it is important to ensure that the fluctuations are not affected by instrumental noise that would cause a systematic error on the amplitudes and spectral indices. For the amplitudes considered in this paper, noise in the magnetic field measurements is insignificant. The noise on the velocity measurements, however, requires a more careful analysis. It is well known that various types of noise, here defined to be anything other than the physical values, can affect turbulence spectra by causing flattening at high frequencies \citep{russell72,johnstone98,podesta06,podesta10d,wicks11,gogoberidze12a}. This flattening can be seen in the velocity spectrum in Figure \ref{fig:spectra}. For the velocity measurements there are three major sources of such noise in the fluctuations (here we do not consider DC offsets): quantization noise due to limited telemetry \citep{russell72}, Poisson noise due to the discrete nature of particles entering the detector \citep{johnstone98}, and aliasing due to discreet sampling \citep{russell72,podesta06}. Here, we consider each of these in turn, adding them to the magnetic field spectrum to investigate their effect, in the spirit of \citet{gogoberidze12a}, but considering physical sources.

\emph{Quantization noise.} Figure \ref{fig:noise}a shows the $E_\mathrm{b}$ and $E_\mathrm{v}$ spectra from Figure \ref{fig:spectra}. To determine the level of quantization in the velocity moments, a PDF of velocity differences at the data resolution ($\approx$ 3 s) was calculated for each component. The PDF shows peaks at multiples of 2.3 km s$^{-1}$ for the GSE x component (the Sunward direction) and multiples of 1.3 km s$^{-1}$ for the other components. These were taken to be the quantization levels, although there is some variation about these values. Figure \ref{fig:noise}b shows $E_\mathrm{v}$ as before along with $E_\mathrm{b}$ normalized to $E_\mathrm{v}$ at $10^{-3}$ Hz, calculated from $\mathbf{b}$ quantized to the same level as the velocity measurements. It can be seen that this quantization does not significantly affect the spectrum and the spectral index remains unchanged in the range $10^{-3}$--$10^{-2}$ Hz. This suggests that quantization noise is not the main source of the flattening of the velocity spectrum at high frequencies.

\emph{Poisson noise.} Each velocity moment is calculated from a measured distribution function, which is typically made up of a few thousand particle counts. This finite number leads to Poisson noise, which causes a white noise spectrum in the velocity moments, with an amplitude that is specific to the instrument design \citep{johnstone98}. To test this effect, Gaussian noise of standard deviation 1.6 km s$^{-1}$ was added to the magnetic field spectrum to cause it to become equal to the velocity spectrum at high frequencies (Figure \ref{fig:noise}c). This represents the maximum amount of Poisson noise (in reality it may be smaller than or equal to this) and is within an order of magnitude of that found by \citet{johnstone98}. It can be seen that even with this maximum amount of Poisson noise, the spectral index remains relatively unchanged at --1.64, compared to --1.66. This suggests that while Poisson noise may account for the high frequency flattening, it is not the cause of the difference in spectral indices between $E_\mathrm{v}$ and $E_\mathrm{b}$.

\emph{Aliasing.} The effect of aliasing on spacecraft power spectra was discussed by \citet{russell72} in the context of magnetic field measurements. A power spectrum $P(f)$ sampled with a Nyquist frequency of $f_\mathrm{N}$ will result in a measured power spectrum of 
\begin{equation}
\label{eq:aliasing}
P'\left(f\right)=P\left(f\right)+\sum\limits_{n=1}^R\left[P\left(2nf_\mathrm{N}-f\right)+P\left(2nf_\mathrm{N}+f\right)\right],
\end{equation}
where $2R+1$ is the ratio of the instrument upper cutoff frequency to the Nyquist frequency. The 3DP instrument on \emph{Wind} generates an onboard velocity moment on each spin of the spacecraft, resulting in a Nyquist frequency of $f_\mathrm{N}= 0.16$ Hz. Since the bulk velocity is larger than the thermal velocity, the solar wind is only seen by the detector for a fraction of the spin period, which can be used to determine the upper cutoff frequency in this interval: the solar wind speed is 440 km s$^{-1}$, the proton thermal speed is 39 km s$^{-1}$, so the 3 sigma solar wind beam width is in view for 6\% of the spin, i.e., 0.18 s, giving an upper cutoff frequency of 2.7 Hz. This gives a ratio of 16.7 and $R\approx 8$. Figure \ref{fig:noise}d shows a --5/3 spectrum together with the aliased version calculated from Equation \ref{eq:aliasing} with these numbers. To construct this, it was assumed that the --5/3 spectrum continues up to 2.7 Hz; this is a worst case scenario, since the ion velocity spectrum has been measured to be steeper than this above 0.2 Hz \citep{safrankova13a}. It can be seen that the aliasing can cause some high frequency flattening but does not significantly alter the spectral index from $10^{-3}$--$10^{-2}$ Hz.

We have shown that these three noise sources are individually not likely to be affecting the velocity measurements in the range $10^{-3}$--$10^{-2}$ Hz, but what about a combination? It is possible to make a model spectrum with the same spectral index as $E_\mathrm{v}$ in this frequency range by taking a --5/3 spectrum, adding white noise at 70 km$^2$s$^{-2}$Hz$^{-1}$ then calculating the aliased spectrum. This, however, produces far more flattening at high frequencies than is seen in the $E_\mathrm{v}$ data (flattening to $>10^{3}$ km$^2$s$^{-2}$Hz$^{-1}$) and results in a non-power law spectrum in $10^{-3}$--$10^{-2}$ Hz, which is not observed. 

To conclude this Appendix, Poisson noise and aliasing are likely to be causing the velocity spectrum to become artificially flat at high frequencies, but not significantly affecting the measurements in the range $10^{-3}$--$10^{-2}$ Hz that is considered in this paper. There are a host of other sources of errors for particle instruments \citep{wuest07}, which are expected to be smaller than those discussed here for the current measurements, but could be considered in a more detailed analysis.

\bibliographystyle{apj}
\bibliography{bibliography}

\end{document}